\documentclass[12pt,showpacs]{revtex4}

\usepackage[english]{babel}
\usepackage{amsmath,amssymb,amsbsy,amstext}
\usepackage[dvips]{graphicx}

\newcommand{\be}{\begin{eqnarray}}
\newcommand{\ee}{\end{eqnarray}}
\newcommand{\bdm}{\begin{displaymath}}
\newcommand{\edm}{\end{displaymath}}
\newcommand{\ds}{\displaystyle}   
\newcommand{\ba}{\begin{array}}
\newcommand{\ea}{\end{array}}
\newcommand{\pa}[1]{\left(#1\right)}
\newcommand{\paq}[1]{\left[#1\right]}
\newcommand{\si}{\sigma}
\newcommand{\Om}{\Omega}
\newcommand{\la}{\lambda}
\newcommand{\sss}{\mathcal{S}}
\newcommand{\tl}[1]{\tilde{#1}}
\newcommand{\snr}{{\rm SNR}}

\begin{document}

\begin{flushright}
\normalsize{gr-qc/0410051}
\end{flushright}

\title{Detecting a stochastic background of gravitational waves by correlating
n detectors}
\author{Orestis Malaspinas and Riccardo Sturani}
\affiliation{Department of Physics, University of Geneva, CH-1211 Geneva,
Switzerland\\
\texttt{e-mail: malaspi9@etu.unige.ch; Riccardo.Sturani@physics.unige.ch}}

\begin{abstract}
We discuss the optimal detection strategy for a stochastic background of 
gravitational waves in the case $n$ detectors are available. In literature so 
far, only two cases have been considered: $2$- and $n$-point correlators. We 
generalize these analysises to $m$-point correlators (with $m<n$) built out of
the $n$ detector signals, obtaining 
the result that the optimal choice is to combine $2$-point correlators.
Correlating $n$ detectors in this optimal way will improve the (suitably 
defined) signal-to-noise ratio with respect to the $n=2$ case by a factor 
equal to the fourth root of $n(n-1)/2$. Finally we give an estimation
of how this could improve the sensitivity for a network of multi-mode spherical
antennas. 
\end{abstract}


\maketitle

\section{Correlation of two detectors}

As it is well known \cite{Christensen:wi}, the sensitivity to a stochastic 
background signal can be greatly enhanced by correlating the output of two 
detectors. To show how this works
it is useful to consider the cross correlation $S_{12}$ 
\cite{Allen:1997ad}
between two detector outputs $S_1$ and $S_2$, defined by
\be \label{random}
S_{12}\equiv \int_{-T/2}^{T/2}dt \int_{-T/2}^{T/2}dt'\ S_1(t)S_2(t')Q(t-t')=
\int_{-\infty}^\infty df \tl S_1^*(f)\tl S_2(f)\tl Q(f)\,,
\ee
where the filter function $Q(t)$ has been introduced. The cross correlation 
$S_{12}$ depends only on the
time difference $t-t'$ as stationarity in both the signal and the noise is 
assumed. In the last equality the Fourier transform of the signal and the
limit $T\to\infty$ have been taken. For any finite $T$, $S_{12}$ is made of
the sum of statistically independent random variable involving $\tl S_1(f)$
and $\tl S_2(f')$, which are correlated only over a frequency range 
$|f-f'|<1/T$.
Thus, as $S_{12}$ is the product of random variables, it is a random variable 
itself and it can be approximated by a Gaussian variable by virtue of 
the central limit theorem, even in the case of narrow band detectors, provided
that $T$ is much larger than the inverse of the bandwidth. 
The same will be true in the case the product of more than two random
variables, that will be considered later.

The outputs of two detectors can be split as $S_{1,2}=s_{1,2}+N_{1,2}$, 
being $s_i$ the physical signal
and $N_i$ the noise. The signal-to-noise ratio for the correlation of the $2$ 
detectors at our disposal (this redundant notation will be useful later, 
where $m$-point correlators out of $n$ detectors will be considered) is given 
by
\be \label{snr}
\paq{\snr(2|2)}^2\equiv\frac{\langle S_{12}\rangle}{\si_{12}}=
\frac{\langle S_{12}\rangle}
{\pa{\langle S^2_{12}\rangle-\langle S_{12}\rangle^2}^{1/2}}=
\frac{\langle s_1s_2\rangle}{\langle N_1^2N_2^2\rangle^{1/2}}\,,
\ee
where $\langle S_{12}\rangle$ and $\si_{12}$ are 
respectively the average and the square root of the variance of the
cross correlation. We have adopted the convention which makes the 
signal-to-noise ratio proportional to the metric perturbation $h$, so that in 
our notation $\snr\propto h$, as in 
\cite{Maggiore:1999vm}, differently from \cite{Allen:1997ad} where 
$\snr\propto h^2$. 
To obtain the last equality in (\ref{snr}) we have made the basic assumptions
that we will never drop throughout this paper: both the signal and the noise 
are Gaussian, they are statistically independent, stationary and with zero 
mean, $N_i\gg s_i$ and finally the noises of different detectors are
completely uncorrelated.

The filter function $Q(t)$ appearing in (\ref{random}) can be freely chosen in 
order to maximise the signal-to-noise ratio. The best choice is obtained in
the standard way
by imposing the functional variation of (\ref{random}) with respect to $Q(t)$ 
equal to zero and solving it for $Q(t)$. To write down the explicit form of the
filter function it is necessary to introduce some further quantity.
The signal can be usefully written as
\be
s_i(t_i,x_i)=\int_{-\infty}^{\infty}df_i\int d\Om_i\tl h_A(f_i,\Om_i)
e^{2\pi if_i\pa{t_i-\Om_ix_i}}F^A(\Omega_i)\,,
\ee
where $\tl h_A$ is the Fourier transform of the metric perturbation with 
polarisation $A$ and $F^A$ is the pattern function of the detectors, which 
encodes the information on its angular sensitivity, $x_i$ the position of the 
$i$-th detector and $\Om_i$ the wave arrival direction.
Given the stochastic nature of the signal, the 2-point correlator (ensemble 
average of the Fourier components) of the metric perturbation can be 
parameterised as
\be \label{sigcor}
\langle\tl h(f_1,\Om_1)\tl h(f_2,\Om_2) \rangle=\delta(f_1+f_2)\frac 1{4\pi}
\delta^2\pa{\Om_1,\Om_2}\frac 1{2}S_h(f_1)\,,
\ee
where the spectral function $S_h$ has been introduced.
Analogously a noise spectral function $S_{N,i}$ for the $i$-th detector can be 
defined through
\be
\langle N_i(f_1)N_j(f_2)\rangle=\delta_{ij}\delta(f_1+f_2)
\frac 12S_{N,i}(f_1)\,.
\ee
The filter function which maximises the signal-to-noise ratio is 
\be \label{optimalq}
\tl Q(f)\propto \frac{S_h(f)\Gamma(f,x_{12})}{S_{N,1}(f)S_{N,2}(f)}\,,
\ee
where the overlap function $\Gamma$ has been introduced. Its definition 
involves the relative distance and orientation of the two detectors
\be
\Gamma(f_i,x_{ab})\equiv\frac 1{4\pi}\int d^2\Om \sum_A F_A^{(a)}(\Om)
F_A^{(b)}(\Om)e^{2\pi i f_i\hat\Om(x_a-x_b)}\,,
\ee
being $F_A^{(a,b)}(\Om)$ the pattern function of the detector at site $a,b$
for a wave coming from direction $\hat\Om$.
Inserting the optimal filter function (\ref{optimalq}) in (\ref{random}) and 
in (\ref{snr}) the explicit form of the signal-to-noise ratio for the 
correlation of two detectors is obtained
\be \label{snr2}
\snr(2,2)=\pa{T\int_{-\infty}^\infty df\ \Gamma^2(f,x_{12})
\frac{S_h^2(f)}{S_{N,1}(f)S_{N,2}(f)}}^{1/4}\,,
\ee
which gains in the case of two identical detectors with respect to the single 
detector case, as it is well known, a factor roughly equal to 
$(T\Delta f)^{1/4}$ multiplied by the overlap function, being $T$ the 
experiment time and $\Delta f$ the bandwidth.

\section{Correlation of n detectors} \label{ndet}

One now might ask what can be gained by the correlation 
of \emph{several} such detectors. A partial answer is obtained by generalizing 
(\ref{snr2}) to the case of $2n$ detectors (the number of
detectors must be even for the correlator not to vanish) \cite{Allen:1997ad}
\renewcommand{\arraystretch}{1.6}
\be \label{snr2n}
\ba{cl}
\snr(2n|2n)&\ds=\frac{\langle S_1\ldots S_{2n}\rangle}
{\pa{\langle N_1^2\ldots N_n^2\rangle}^{1/2}} = \\
&\ds= T^{1/4}\pa{\int df_1\ldots\int
df_n\frac{\pa{\prod_{i=1}^nS_h(f_i)\Gamma(f_i,x_{i,n+i})}^2
+\rm{perm}}{\prod_{i=1}^{n}S_{N,i}(f_i)S_{N,i+n}(-f_i)}}^{1/4n},
\ea
\ee
\renewcommand{\arraystretch}{1}
where in our notation $\snr(i|j)$ is the signal-to-noise ratio 
given by $i$-point correlators taken out of $j$ detectors.
To obtain (\ref{snr2n}) the explicit form of the optimal filter function
\cite{Allen:1997ad}, which is determined up to an arbitrary constant,
\renewcommand{\arraystretch}{1.6}
\be
\ba{l}
\tl Q(f_2,\ldots,f_{2n})\propto\\
\qquad
\ds\frac{S_h(-f_{n+1})\Gamma(-f_{n+1},x_{1,n+1})
\paq{\prod_{i=2}^{n}S_h(f_i)\Gamma(f_i,x_{i,n+i})\delta(f_{n+i}+f_i)}+
\rm{perm}} {S_{N,1}(-f_{n+1})\prod_{i=2}^{2n}S_{N,i}(f_i)}
\ea
\ee
\renewcommand{\arraystretch}{1}
has been used.
The permutations come from all the different pairings of $2n$ 
signals: as the detector output is Gaussian its $n$-point correlator
can be computed from the product of 2-point ones. In $\snr(2n|2n)$ we 
indicated only the terms with leading behaviour in $T$ for large $T$, which are 
$(2n-1)!!$. Eq. (\ref{snr2n}) can be rewritten schematically as
\be
\paq{\snr(2n|2n)}^{4n}=\prod_{i=1}^n \paq{\snr(i,n+i|2)}^4+\rm{perm}\,,
\ee
i.e. it is the sum of permutations of products of $2$-point correlators.

We now show that there exists a \emph{better} way to treat data obtained
from $2n$ detectors, as out of $2n$ detectors, $2m$-correlators can be 
considered, for any $m<n$.
For $m=1$  we can follow the analysis of \cite{Christensen:wi} or 
\cite{Allen:1997ad} and consider all the possible pairs taken out of $2n$ 
detectors. For each detector pair a mean value and a variance can be defined 
as usual
\be
\bar S_{ij}\equiv\langle S_{ij}\rangle=\bar S_2\qquad 
\si^2_{ij}=\langle S^2_{ij}\rangle-\bar S^2_{ij}\,,
\ee
where the optimal filter function has been normalized so to make the 
theoretical mean $\langle S_{ij}\rangle=\bar S_2$ equal for every pair.
A $\snr(i,j|2)$ of the type (\ref{snr2}) can thus be assigned to each pair
\be \label{snrij}
\paq{\snr(i,j|2)}^2=\frac{\bar S_{ij}}{\si_{ij}}\,.
\ee
The best way to gather the information from all the pairings is to take a 
weighted average with weights $\la_{ij}$
\be \label{comb}
\sss_2\equiv\frac{\sum_{i<j} \la_{ij} S_{ij}}{\sum_{i<j}\la_{ij}}\,,
\ee
whose variance is
\bdm
\sigma^2_{\sss_2}\equiv\langle\sss^2_2\rangle-\langle\sss_2\rangle^2=
\frac{\sum_{i<j}\la_{ij}^2\sigma^2_{ij}}{\pa{\sum_{i<j}\la_{ij}}^2}\,,
\edm
which is justified by large noise approximation we are using, 
that allows to neglect non diagonal terms like $\si_{ij}\si_{kl}$ (for
$\{i,j\}\neq\{k,l\}$) compared to $\si^2_{ij}$.
The signal-to-noise ratio obtained by combining in pairs the $2n$ detector 
outputs in this way is given by 
\be
\paq{\snr(2|2n)}^4=\frac{\langle\sss_2\rangle^2}{\si^2_{\sss_2}}=
\frac{\pa{\sum_{i<j}\la_{ij}\bar S_{ij}}^2}{\sum_{i<j}\la^2_{ij}\si^2_{ij}}\,.
\ee
The best signal-to-noise ratio is obtained by choosing 
$\la_{ij}\propto \si_{ij}^{-2}$ (which correspond to weighing less the more
noisy data) and it is 
\be \label{snr2con2n}
\paq{\snr(2|n)}^4=\sum_{i<j}\frac {\bar S_2^2}{\si_{ij}^2}=
\sum_{i<j}\paq{\snr(i,j|2)}^4\,,
\ee
where we have dropped the unnecessary hypotheses of the number of 
detectors being even.
The optimal signal-to-noise ratio is thus given by the sum of terms like 
(\ref{snr2}) (to the fourth power); note that we recover the time dependence
of (\ref{snr2n}): $\snr(2|n)\propto T^{1/4}$ \cite{Christensen:wi}. 
For $n$ detectors with equal noise level, data collection time and overlap 
functions, we have 
\be \label{ndep}
\snr(2|n)\propto [n(n-1)]^{1/4}\,.
\ee

We now generalize the analysis of the combination of $2$-points correlators
to the case of $2m$-point correlators.
Analogously to (\ref{comb}) we can define a linear combination $\sss_{2m}$ of 
the $(n)!/[(2m)!(n-2m)!]$ $2m$-point correlators $S_{i_1\ldots i_{2m}}$
that is possible to build out of $n$ detectors.
Defining a signal-to-noise ratio of the type (\ref{snr2n})
\bdm
\paq{\snr(i_1\ldots i_{2m}|2m)}^{2m}\equiv 
\frac{\langle S_{i_1\ldots i_{2m}}\rangle}{\si_{i_1\ldots i_{2m}}}=
\frac{\bar S_{2m}}{\si_{i_1\ldots i_{2m}}}\,,
\edm
as a natural generalization of (\ref{snrij}), we are led to consider the 
combination of the $2m$-correlators analogous to (\ref{comb}) 
\bdm
\sss_{2m}\equiv \frac{\sum_{i_1<\ldots <i_{2m}}
\la_{i_1\ldots i_{2m}} \bar S_{2m}}
{\sum_{i_1<\ldots<i_{2m}}\la_{i_1\ldots i_{2m}}}\,, \qquad
\si^2_{\sss_{2m}}\equiv \frac{\sum_{i_1<\ldots <i_{2m}}
\la_{i_1\ldots i_{2m}}^2 \si^2_{i_1\ldots i_{2m}}}
{\pa{\sum_{i_1<\ldots<i_{2m}}\la_{i_1\ldots i_{2m}}}^2}\,,
\edm
(with $i_k\in \{1\ldots 2n\}$) so that the signal-to-noise ratio for 
$2m$-correlators can be written, for the optimal choice of weights
$\la_{i_1\ldots i_{2m}}\propto \si^{-2}_{i_1\ldots i_{2m}}$, as 
\be \label{snr2m2n}
\paq{\snr(2m|2n)}^{4m}\equiv\frac{\langle\sss_{2m}\rangle^2}{\si^2_{\sss_{2m}}}
=\sum_{i_1<\ldots <i_{2m}}\frac{\langle S_{i_1\ldots i_{2m}}\rangle^2}
{\si^2_{i_1\ldots i_{2m}}}=
\sum_{i_1<\ldots <i_{2m}}\paq{\snr(i_1\ldots i_{2m}|2m)}^{4m}\,.
\ee
Each of the terms in the sum in the  most rhs of (\ref{snr2m2n}) is on its own 
the sum of $(2m-1)!!$ terms as shown in (\ref{snr2n}).

For equal noises, observation times and overlap functions the scaling of the 
signal-to-noise ratio with respect to the number of detectors $n$, and with
the order of the correlator $2m$, is given by
\be
\snr(2m|n)\propto \paq{(2m-1)!!\times\pa{\ba{c}n \\ 2m\ea}}^{\frac 1{4m}}\,,
\ee 
where the first factor comes from the number of contribution in each 
$\snr(i_1\ldots i_{2m}|2m)$ and the binomial coefficient from the possible
choices of $2m$-ple out of $n$ detectors.

For any fixed $n$ the maximum is obtained always for $m=1$ implying that the 
optimal signal-to-noise ratio is obtained by combining the detectors in pairs 
as in (\ref{snr2con2n}). In particular for a network made of a large 
number of detectors the signal-to-noise is expected to scale with the square 
root of the number of detectors as in (\ref{ndep}).

\section{Stochastic background}

To parameterise conveniently the detector sensitivity to a stochastic 
background it is useful to introduce the normalized spectral energy density of 
gravitational waves $\Om_{gw}(f)$ defined as follows
\be
\Om_{gw}(f)\equiv \frac 1{\rho_c}\frac{d\rho_{gw}(f)}{d\ln f}\,,
\ee
where $\rho_c=(3h^2_{100}H_0^2)/(8\pi G_N)$ is the critical energy density of 
the Universe ($h_{100}H_0$ is the Hubble constant and 
$H_0\equiv 100\rm{Mpc^{-1}km/s}$) and $\rho_{gw}$ 
is the Fourier transform of the energy density in gravitational wave. In 
terms of the spectral function $S_h$ introduced in (\ref{sigcor}) $\rho_{gw}$ 
can be written as
\be
\frac{d\rho_{gw}(f)}{d\ln f}=\frac{\pi}{2G_N}f^3 S_h(f)\,.
\ee
Using this formula it is possible to rewrite (\ref{snr2}) in terms of 
$\Om_{gw}$ 
\be
\paq{\snr(2|2)}^2=\sqrt{2T}\frac{3 h^2_{100}H_0^2}{4\pi^2}
\paq{\int_0^\infty df\frac{\Gamma^2(f,x_{1,2})\Om_{gw}^2(f)}
{f^6S_{N,1}S_{N,2}}}^{1/2}\,,
\ee 
which can be used in (\ref{snr2con2n}) to express the $\snr(2|n)$ as a 
function of $\Om_{gw}$. The SNR necessary to claim detection can be computed 
once a \emph{false alarm rate} $\alpha$ and a \emph{false dismissal 
rate} $1-\gamma$ are specified ($\gamma$ is called \emph{detection rate}), 
according to the Neymann-Pearson detection 
criterion (see \cite{Allen:1997ad} for the definition of false alarm and false 
dismissal rates). A natural choice is $\alpha=5\%$ and $\gamma=95\%$.
 
Even if it is useful in principle to correlate as many detectors as possible,
in practice the number of high sensitivity detectors available with non 
negligible overlap function is not too large.
For instance in \cite{Allen:1997ad} it is shown that with the current light 
interferometers it is possible to detect a stochastic background of 
gravitational waves provided that their normalized energy density satisfies the
bound
\be \label{LLVT}
h^2_{100}\Om_{gw}^{95\%,5\%}\gtrsim 6.5\cdot 10^{-6}\,,
\ee
for $\gamma=95\%$ and $\alpha=5\%$, constant $\Om_{gw}$ and an observation 
period of 4 months. This bound is obtained by correlating the two LIGO's, 
VIRGO and GEO600, but a numerically similar one is obtained by correlating 
just the two LIGO's.

From the phenomenological point of view it is interesting to note that the 
total gravitational wave energy stored in a stochastic background cannot 
exceed the bound
\be \label{nucl}
\Om_{gw}\lesssim 6\cdot 10^{-6}\,,
\ee
surprisingly numerically close to (\ref{LLVT}), otherwise the Universe would 
expand too rapidly in the epoch of primordial nucleosynthesis, 
thus spoiling the beautiful agreement between theory and observation of 
the primordial abundance of light elements. 
The nucleosynthesis epoch corresponds to ten seconds after the Big 
Bang or to a temperature of the order of few MeV.

This limit does not apply to a background of gravitational waves produced 
\emph{after} the nucleosynthesis epoch. There exist indeed astrophysical 
sources which may produce a continuous stochastic signal in the 
phenomenologically interesting frequency range. For instance rapidly 
rotating young neutron stars could be the source of gravitational 
radiation with an amplitude of $h^2_{100}\Om_{gw}\sim 10^{-8}$ for the 
frequency range $0.7- 1$ kHz \cite{Ferrari:1998jf}.

A system which is able of producing gravitational wave to an observable level 
is represented by cusps and kinks in cosmic string network 
\cite{Damour:2004kw}. For some range of the parameters entering 
the description of the cosmic strings the energy of gravitational wave 
$\Om_{gw}$ could be as high as $10^{-6}$ and it could be produced either
before or after the nucleosynthesis epoch.

\section{Network of antennas}

We have shown in sec.~\ref{ndet} that with $n$ detectors available, the best 
strategy to detect a stochastic background is to correlate pairs of them, and 
in case noise levels and overlap functions have equal values the 
signal-to-noise ratio increases as $n^{1/2}$ for large $n$.

Let us now turn our attention to a case in which it is important to correlate 
a \emph{large} number of detectors to increase the detector sensitivity. 

An interesting case can be realized by multi-mode detectors like spherical 
antennas \cite{Zhou:1995en} as they have \emph{five} quadrupolar modes which 
couple to a gravitational wave with generic incident direction. The
correlation of such modes among several antennas can be considered, thus 
increasing the number of effective detectors available.

\begin{figure}
\centering
\includegraphics[width=.6\linewidth]{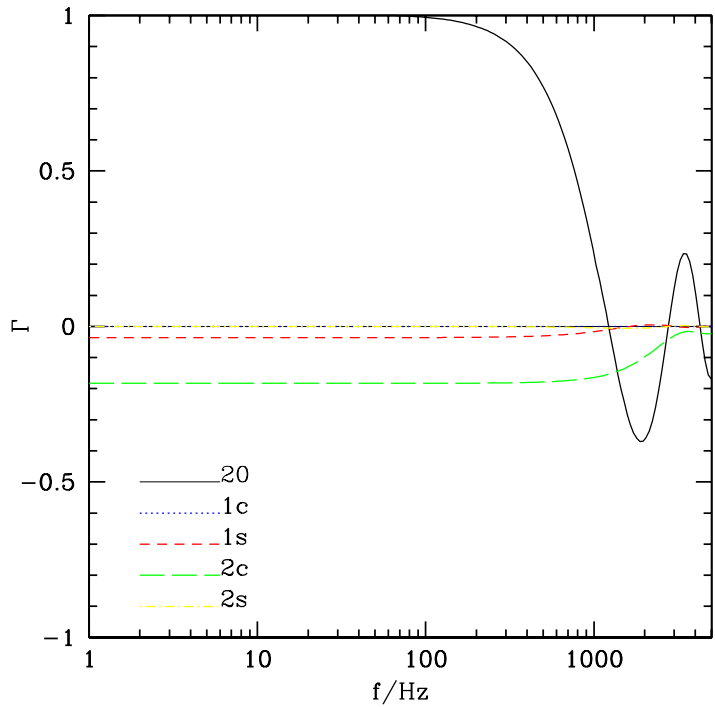}
\caption{Overlap reduction function, as a function of the frequency, between 
the $m=0$ mode of a sphere
at lat.~$52.2^o$ N, long.~$4.5^o$ E, (quantisation axis chosen along the local 
meridian) and the five quadrupolar modes (labelled $20$, $1c$, $1s$, $2c$, $2s$
after \cite{Zhou:1995en}) of a second sphere located at 
$100$ Km (lat.~$52.2^o$ N, long.~$6.0^o$ E, quantization axis along the 
local meridian).}
\label{overlap00}
\end{figure}

Anyway it has to be considered that modes of the same sphere cannot be 
correlated among themselves, as their noises \emph{are} correlated 
and most of the mode pairs have negligible values of the overlap functions. 
Fig.~\ref{overlap00} shows
the overlap reduction function for different pairs of modes.
Denoting by $m$ the integer number labelling different quadrupolar modes 
($-2\leq m\leq 2$), the five overlap functions in the figure are 
obtained by correlating the $m=0$ mode of a sphere and the five modes
of a second sphere located at $100$ Km (the quantisation axes have been 
chosen parallel to each other in order to maximise the sum of the 
overlap functions).

\begin{figure}
\centering
\includegraphics[width=.6\linewidth]{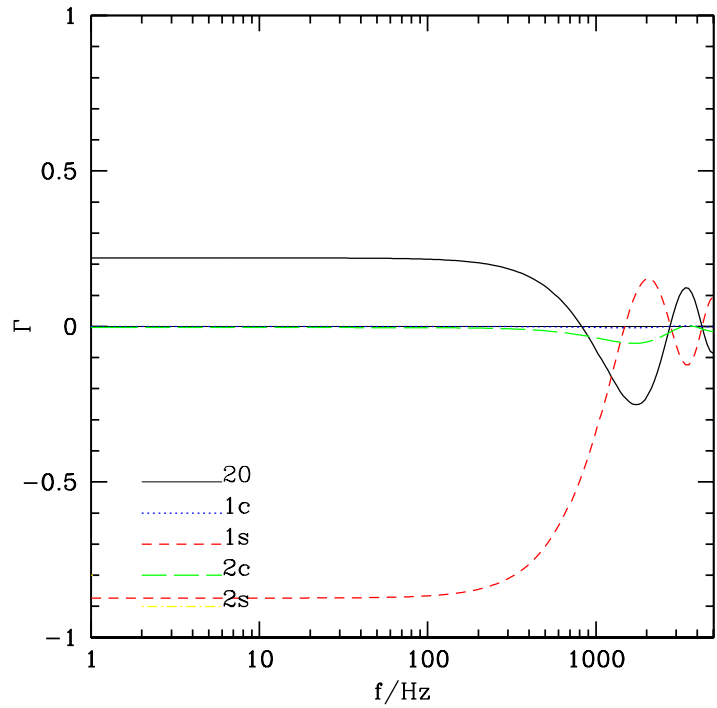}
\caption{Overlap reduction function,as a function of the frequency, between 
the $m=0$ mode of a sphere
at lat.~$52.2^o$ N, long.~$4.5^o$ E, (quantization axis along the local 
meridian) and the five quadrupolar modes (labelled $20$, $1c$, $1s$, $2c$, $2s$
after \cite{Zhou:1995en}) of a second sphere located at 
$100$ Km (lat.~$52.2^o$ N, long.~$6.0^o$ E, quantization axis at $45^o$ 
with respect to the local meridian).}
\label{overlap45}
\end{figure}

This makes the statistics a bit different than in (\ref{ndep}), so that for a
set of $n$ spherical antennas the analogous of (\ref{snr2con2n}) is
\be \label{nsfcor}
\paq{\snr_{sphere}(2|n)}^4=\sum_{i<j\leq n}
\sum_{m,m'=1}^5\paq{\snr(i,m;j,m'|2)}^4\,,
\ee
where the indices $i,j$ run over different detectors and $m,m'$ run over the 
quadrupolar modes. If all the modes are equally noisy it is clear from 
fig.~\ref{overlap00} that every mode of the first sphere can be effectively 
correlated with just one or two modes of the second sphere, as it has almost 
vanishing overlap with the others.
The situation does not improve by choosing a different orientations for the 
quantization axis of different spheres, as an example for an angle of $45$ 
degrees shows in fig.~\ref{overlap45}.

At the moment the only operating spherical antenna is miniGRAIL (and another
one will soon start working, the Brazilian Gravitational Wave Detector Mario 
Schenberg \cite{Aguiar:2005tp}), thus to obtain some numbers let us consider 
the present noise spectral function of miniGRAIL \cite{minigrail}, whose 
diameter is $68$ cm. It can presently reach a \emph{strain sensitivity} $h_c$ 
of about $h_c\equiv \sqrt{S_h}\sim 10^{-20}Hz^{-1/2}$, it has a resonant 
frequency of $2.9$kHz and a bandwidth of about $230$ Hz. For this figures one 
obtains, for a single pair of detectors 
\be
h^2_{100}\Om^{95\%,5\%}_{gw}=\frac 1{\sqrt 5}\frac 1{\sqrt{2T}}
\frac{4\pi^2}{3H_0^2}
\paq{\int_0^\infty df\frac{\Gamma^2(f,x_{12})}{S_N^2(f)}}^{-1/2}\sim 10
\ee 
for an observation time of 1 year. 
Using a set of identical spheres eq. (\ref{nsfcor}) can be explicited as
\be \label{sphe}
\paq{\snr_{sphere}(2|n)}^4=2T\pa{\frac{3h^2_{100}H_0^2}{4\pi^2}}^2
\int_0^{\infty}df\frac{\Om_{gw}^2(f)\sum _{i<j}\sum_{m,m'}
\Gamma^2(f,{x_{i,j}})}{f^6 S_N^2(f)}\,,
\ee
where $S_N(f)$ is the common noise spectral function and $\Gamma(f,x_{i,j})$
is the overlap function for the detector pair $i-j$, which is understood to 
depend also on $m$ and $m'$. 
The situation can be considerably improved by using larger
spheres, with a consequent lower resonant frequency. We can estimate for
instance that slightly improving the sensitivity to 
$h_c\simeq 10^{-21}Hz^{-1/2}$, a resonant frequency at, say, $300$ Hz (and 
bandwidth $100$ Hz) one can reach
\be \label{estss}
h^2_{100}\Om^{95\%,5\%}_{gw}\simeq 4 \cdot 10^{-4}\times
\paq{5\cdot\frac{n(n-1)}2}^{-1/2}\,,
\ee
which is obtained by inverting (\ref{sphe}) for a constant $\Om_{gw}$,
thus allowing to obtain $\Om_{gw}\simeq 3\cdot 10^{-5}$ in the experiment 
bandwidth for a set of $n=10$ detectors. We note that it is important to have
a not too high resonant frequency for detector correlation, as overlap
functions go to zero at $f\gtrsim 1/\pi L$, being $L$ the detector separation.
This still far from light interferometers, see eq.(\ref{LLVT}) and 
the phenomenological bound given by eq.(\ref{nucl}), but the effect of the 
multiple 
correlation is quantitatively important, so it is not excluded that once 
higher sensitivity will be achieved the correlation effect will be crucial for
detection.

\begin{figure}
\centering
\includegraphics[width=.6\linewidth]{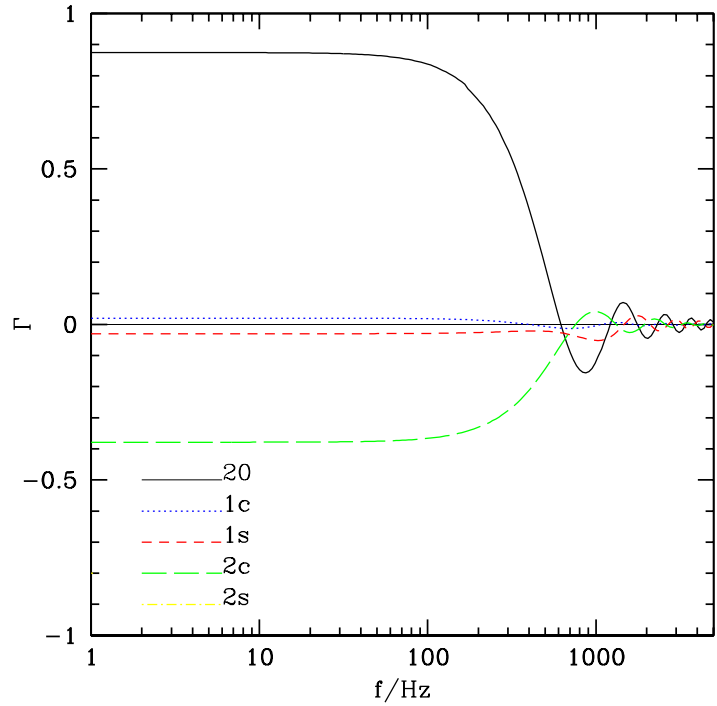}
\caption{Overlap reduction function between VIRGO (lat.~$43.6^o$ N, 
long.~$10.5^o$ E, azimuth $160.6^o$ E) and the five 
quadrupolar modes of a hypothetical sphere located in Rome 
(lat.~$41.8^o$ N, long.~$12.6^o$ E, quantization axis at $-20^o$ E).}
\label{overlapvs}
\end{figure}

The mechanism of sensitivity enhancement actually will not work if a sphere
is correlated with an interferometer as only one mode of the sphere
can effectively be correlated with an interferometer. This can be seen in
fig.~\ref{overlapvs}, which shows the overlap function between VIRGO and the 
five quadrupolar modes of a hypothetical sphere placed in Rome.
A sphere like one with the characteristics leading to (\ref{estss}), has a 
narrower bandwidth than an interferometer, but similar sensitivity in its
bandwidth, so correlating a sphere with an interferometer would lead to 
a result equal to (\ref{estss}) but for the factor in square brackets, as one
cannot take advantage of the correlation of several modes of the same sphere:
correlation with an interferometer would just add one more single-mode 
detector.

\section{Conclusion}

We have analyzed the utility of considering multiple detector correlators to 
detect a stochastic background of gravitational waves. The main result of the
paper is the demonstration that the best way to correlate the outputs of
different detectors is in pairs, no matter how large is the number of 
detectors, instead of taking $m$-correlators with $m>2$. 
Finally as a potentially interesting application of this result we applied 
this strategy to a set of identical spheres, showing that correlation of 
several pairs of detectors is important in increasing the sensitivity.

\section*{Acknowledgements}

We thank Florian Dubath, Stefano Foffa and Alice Gasparini and Michele
Maggiore for useful discussions. The work of R. S. is supported by the Fonds 
National Suisse.


\section*{References}

\begin{thebibliography}{10}

\bibitem{Christensen:wi}
N.~Christensen,
Phys.\ Rev.\ D {\bf 46} (1992) 5250.

\bibitem{Allen:1997ad}
B.~Allen and J.~D.~Romano,
Phys.\ Rev.\ D {\bf 59} (1999) 102001

\bibitem{Maggiore:1999vm}
M.~Maggiore,
Phys.\ Rept.\  {\bf 331} (2000) 283

\bibitem{Ferrari:1998jf}
V.~Ferrari, S.~Matarrese and R.~Schneider,
Mon.\ Not.\ Roy.\ Astron.\ Soc.\  {\bf 303} (1999) 258

\bibitem{Damour:2004kw}
T.~Damour and A.~Vilenkin,
Phys.\ Rev.\ D {\bf 71} (2005) 063510

\bibitem{Zhou:1995en}
C.~Z.~Zhou and P.~F.~Michelson,
Phys.\ Rev.\ D {\bf 51} (1995) 2517.

\bibitem{minigrail}
A. de Waard et al.,''Progress report 2004'',
Proceedings of the 5th International LISA Symposium, 
European Space Research and Technology Centre (ESTEC), Noordwijk, 
The Netherlands, Classical and Quantum Gravity (2005) 

\bibitem{Aguiar:2005tp}
O.~D.~Aguiar {\it et al.},
Class.\ Quant.\ Grav.\  {\bf 22} (2005) S209.

\end{thebibliography}
\end{document}